# ARTICLE

# Evolution and control of the phase competition morphology in a manganite film

Haibiao Zhou[1,2], Lingfei Wang[2], Yubin Hou[1], Zhen Huang[2], Qingyou Lu[1,2,3] & Wenbin Wu[1,2,3]

The competition among different phases in perovskite manganites is pronounced since their energies are very close under the interplay of charge, spin, orbital and lattice degrees of freedom. To reveal the roles of underlying interactions, many efforts have been devoted towards directly imaging phase transitions at microscopic scales. Here we show images of the charge-ordered insulator (COI) phase transition from a pure ferromagnetic metal with reducing field or increasing temperature in a strained phase-separated manganite film, using a home-built magnetic force microscope. Compared with the COI melting transition, this reverse transition is sharp, cooperative and martensitic-like with astonishingly unique yet diverse morphologies. The COI domains show variable-dimensional growth at different temperatures and their distribution can illustrate the delicate balance of the underlying interactions in manganites. Our findings also display how phase domain engineering is possible and how the phase competition can be tuned in a controllable manner.

[1] High Magnetic Field Laboratory, Chinese Academy of Sciences and University of Science and Technology of China, Hefei 230031, China. [2] Hefei National Laboratory for Physical Sciences at Microscale, University of Science and Technology of China, Hefei 230026, China. [3] Collaborative Innovation Center of Advanced Microstructure, Nanjing University, Nanjing 210093, China. Correspondence and requests for materials should be addressed to Q.L. (email: qxl@ustc.edu.cn) or to W.W. (email: wuwb@ustc.edu.cn).

Phase coexistence even in a single-crystal, or phase separation (PS), has been found ubiquitous in strongly correlated electron systems[1,2]. It is most prominent in manganites and considered as a prerequisite for the colossal magnetoresistance[1]. Growing efforts have been devoted to investigating PS in manganites including ultrafast processes using pump-probe methods[3,4]. However, this type of method can usually only provide time-domain information, and to probe the phase inhomogeneity, techniques that offer spatial resolution are indispensable. In fact, some progresses have been made[5–13], such as a direct observation of percolation in $La_{3/8-y}Pr_yCa_{3/8}MnO_3$ (LPCMO)[5,11], although most of them are performed without the presence of a magnetic field, a key parameter capable of tuning phase competition (PC). Recently, scanning probe microscopies under high magnetic fields, such as magnetic force microscopy[7] (MFM) and microwave impedance microscopy[10], have been developed to explore PS in manganites, in particular, the melting transition from an antiferromagnetic charge-ordered insulator (COI) ground state to a ferromagnetic metal (FMM) state under high fields, revealed the important roles of crystal defects[7] and epitaxial strain[10]. But even so, the reversed phase transition to the COI phase from a fully saturated FMM on decreasing the external magnetic field, though observed broadly through transport measurements of a series of manganite systems with a COI ground state, such as $Nd_{0.5}Sr_{0.5}MnO_3$ (ref. 14), $Pr_{0.5}Ca_{0.5}MnO_3$ (ref. 15) and LPCMO[16] has surpringly never been imaged at microscopic scales, possibly due to a larger melting field needed[14,15]. We label this field-driven reversed transition of COI as FR for convenience. Surely, this process is of the equal importance as its melting counterpart and should probably complement new information for the further understanding of phase dynamics in manganites. Furthermore, quite different PC morphologies were observed for different samples in the melting process, such as patch-like phase domains in LPCMO[5,7,12] and cross-like FMM filaments in $Nd_{0.5}Sr_{0.5}MnO_3$ films[10]. Therefore, a question may arise as to whether the PC patterns are sample dependent or rely on the stage of PS evolution.

Herein, in addition to the melting process, the COI phase transitions from a magnetic-field-induced saturated FMM state as functions of magnetic field ($H$) and temperature ($T$) in a single sample are imaged using a home-built MFM, which features a home-designed scan head housed in a 20 T superconducting magnet (see Methods section and ref. 17 for details). The sample is a 55-nm annealed $La_{0.67}Ca_{0.33}MnO_3$ (LCMO) film grown on a (001)-oriented NdGaO$_3$ (NGO) substrate. Bulk LCMO at this doping level shows only an FMM transition, while the films can behave differently, for example, a COI ground state has been induced in LCMO/LaAlO$_3$ films[18]. Very recently, we found that films grown on NGO(001) can show dramatically different PC behaviour[19–21]. By imaging these transitions systematically, we found that the phase domain behaviour during FR is unexpected and vastly different from the melting transition. These results would substantially increase our understanding of PC in manganites, which may also extend to other similar materials.

## Results
**Transport measurements**. The annealed LCMO/NGO(001) film, characterized by an anisotropic epitaxial strain (Fig. 1a inset), shows two successive but incomplete first-order phase transitions, that is, FMM-to-COI at $T_{AFI} = 250$ K and COI-to-FMM at $T_C' = 118$ K below the Curie temperature ($T_C = 265$ K), as revealed by the temperature-dependent resistivity ($\rho$–$T$) curves (Fig. 1a). The large hysteresis in temperature implies that the system is phase separated. The PS regime below $T_{AFI}$ can be further divided into the COI-dominated PS (COI-PS), the FMM-dominated PS (FMM-PS) and the frozen states[7,19,22]. More detailed transport measurements can be found in Supplementary Fig. 1 and ref. 19, where the $T$–$H$ phase diagram of the film can be seen as a combination of two prototype systems, $La_{0.5}Ca_{0.5}MnO_3$ and LPCMO. The differences between the COI-PS, FMM-PS and frozen states are further revealed by the $\rho$-$H$ curves shown in Fig. 1b. After the COI phase melts (path 1, Fig. 1b), on decreasing the field the COI phase reappears as evidenced by the upturn in $\rho$ in the COI-PS state (path 2, Fig. 1b), which has been labelled as FR. Note that at 150 K the 'overshot' in resistivity is observed[23]. However, this transformation does not occur for the FMM-PS or frozen states (10 K, Fig. 1b). The transition temperature $T_C'$ between COI-PS and FMM-PS states can be extracted from the $\rho$–$T$ curves undergoing a zero-field warming after a field cooling process. An abrupt increase in $\rho$ occurs at $T_C'$ (Fig. 1c) and we label this temperature-driven reappearance of COI phase as TR. It should be noted that the transitions between the FMM and COI phases are accompanied by structural transitions since the FMM phase has an orthorhombic structure with weak Jahn–Teller distortions and the COI phase has a monoclinic-like structure with strong Jahn–Teller distortions[22]. Thus, for this type of film, in addition to the anisotropic epitaxial strain stemming from the substrate there also exist accommodation strains due to a structural mismatch at the interfaces between the FMM and COI phase domains[9,11,22,24–26]. The melting, FR and TR occurring in the same sample makes it a perfect platform to explore the detailed microscopic PC behaviour in manganites.

**MFM images**. Figure 2 shows MFM images where the field is swept up to $\mu_0H = 4.0$ T at 150 K. At this temperature the sample is in the COI-PS state (Fig. 1b). The film is first cooled from room temperature to 150 K at 0 T, where the MFM image (Fig. 2a) shows a cloud-like mixture of dark and bright regions. As the field is increased to 0.4 T, the contrast of the MFM image (Fig. 2b) is enhanced owing to the aligned magnetic moments of the existing FMM phase (see Methods section for the magnetic properties of the tip and the image contrast analysis). Dramatic changes occur at 1.8 T (Fig. 2c), where the contrast becomes very clear and dense stripes form along the orthorhombic $a$ axis, as reflected by the fast Fourier transformation pattern (Fig. 2c inset). The separated dark and bright regions are $\sim 0.2\,\mu m$ wide and correspond to the FMM and COI phases, respectively. The FMM paths along the $a$ axis suggest the presence of anisotropic percolation and resistance[27] (Supplementary Fig. 2). At 2.7 T (Fig. 2d), the dark FMM paths begin to expand, whereas the bright COI stripes shrink and become segmented and broken into isolated droplets in still higher fields (Fig. 2e,f). At 4.0 T (Fig. 2g), the sample is fully FMM with weak contrast arising from the topography (see Supplementary Fig. 3 for details). The single remaining bright spot is probably a defect, which serves as a field-independent marker.

Much more drastic behaviour is seen as we reduce the field from the fully saturated FMM state. This FR (path 2, Fig. 1b) has been observed in the transport properties in a variety of manganites[14,15]. However, to the best of our knowledge, this process has never been explored microscopically and it turns out to be very distinct here. In the beginning, no apparent changes can be found until $\mu_0H = 1.5$ T (Fig. 2h), where a single giant COI stripe elongated along the $a$ axis appears. This field is larger than the FR onset field in the $\rho$-$H$ measurement (0.96 T, path 2 in Fig. 1b), and the width of the COI stripe ($\sim 0.4\,\mu m$) is much larger than that observed in the above melting process. As the field is further lowered, more giant stripes appear. Surprisingly, some of them show striking discontinuities along the slow scan direction (top to bottom) marked by the white arrows (Fig. 2j,k), meaning that these stripes grow rapidly: a stripe that has not yet begun to grow when the tip scans a certain line may become fully grown when the tip scans the next line. According to our measurements, such stripes can be fully grown within 60 ms (see Supplementary Fig. 4 for calculation details). Therefore, the reappearance of COI can be classified to undergo a microscopic 'umklapp transformation'[25], demonstrating that the emergence of the COI domains does not occur through atomic diffusion, but rather is mediated by the cooperative distortions of $MnO_6$ octahedra. This is consistent with the martensitic-like transformation mechanism proposed for the strain state transitions[25,28,29], and the zigzag COI boundaries also suggest that the growth is impeded by the accommodation strain[9,24].

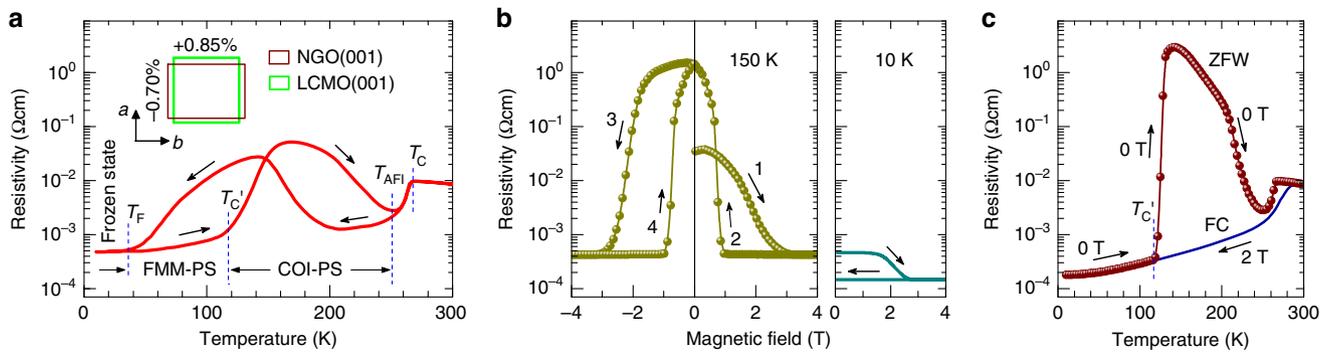

**Figure 1 | Magnetotransport measurements of the LCMO/NGO(001) film along the *b* axis. (a)** The zero-field $\rho$–$T$ curve. $T_C$, $T_C'$ and $T_{AFI}$ denote the FMM transition temperatures at 265 K and 118 K, and the onset COI transition temperature at 250 K, respectively. The inset illustrates the anisotropic strain supplied by the substrate with a higher orthorhombicity. In Pbnm notation the lattice parameters of bulk LCMO (B) and the substrate (S) are $a_B = 5.4717$ Å, $b_B = 5.4569$ Å, $c_B = 7.7112$ Å and $a_S = 5.4332$ Å, $b_S = 5.5034$ Å, $c_S = 7.7155$ Å, respectively. Full epitaxial strain state can be induced by the annealing, as shown by the detailed structural characterizations in ref. 21. **(b)** The $\rho$-$H$ curves at 150 and 10 K. **(c)** The resistivity during field cooling (FC) at 2.0 T (line) followed by zero-field warming (ZFW) (symbol and line), where a sharp jump is observed at $T_C'$. In **b** and **c**, the magnetic field is applied perpendicular to the film plane. All the arrows represent the time sequence.

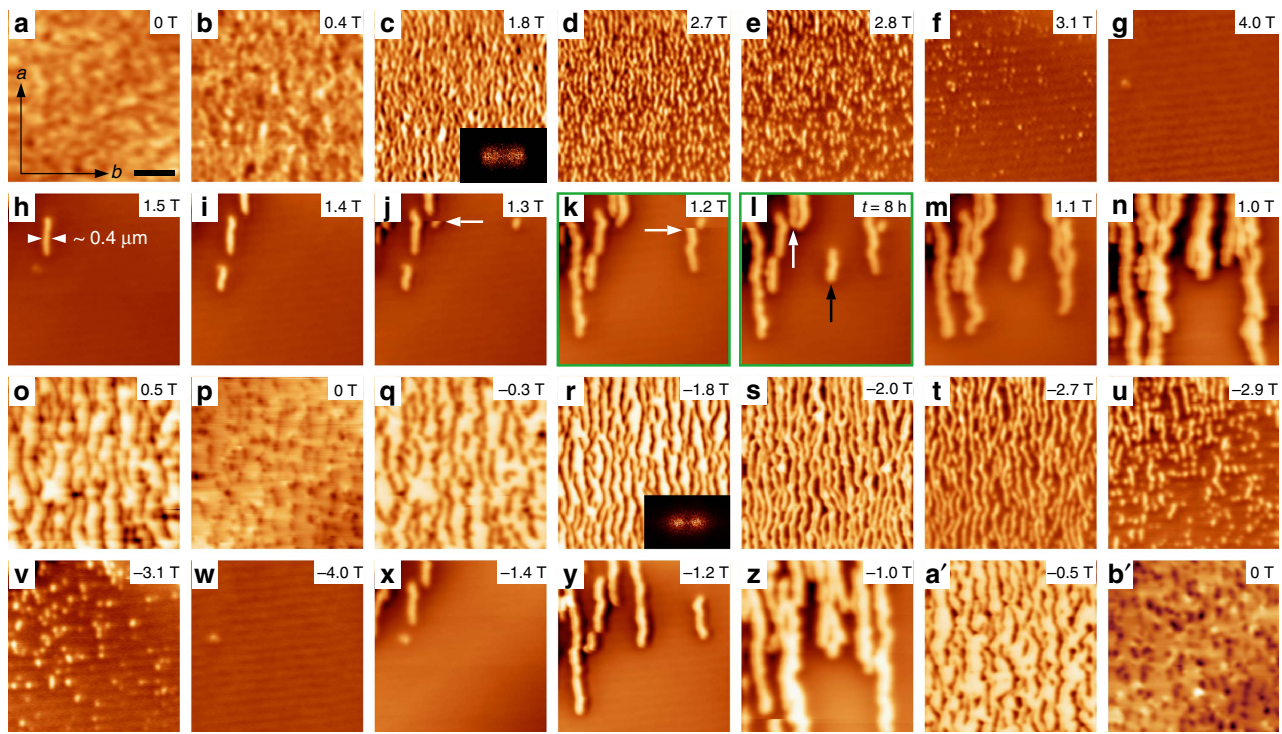

**Figure 2 | MFM images taken at the same location during field sweep at 150 K. (a–g)**, **(h–p)**, **(q–w)** and **(x–b')** are four image series, which correspond to the four numbered paths in the transport measurements. The sample is first cooled from room temperature to 150 K and then the magnetic field is applied perpendicular to the film plane. The scale bar in **a** is 2 μm and the scanned areas are 8.5 μm × 8.5 μm for all images. The colour scales for **a–b'** are 0.6, 0.6, 0.6, 0.4, 0.4, 0.4, 0.4, 1.2, 1.5, 2, 2, 2.5, 3, 0.4, 0.3, 1, 0.6, 2.0, 0.8, 0.6, 0.3, 0.3, 1.5, 2.5, 2.5, 2.5 and 0.3 Hz, respectively. The insets in **c** and **r** are the fast Fourier transformation patterns of 14 μm$^{-1}$ × 9.0 μm$^{-1}$, exhibiting evident dumbbell-like density distribution in both cases. The white arrows in **j** and **k** indicate sharp discontinuities, and the dark and white arrows in **l** denote the growing of existing and new COI domains at a fixed field 1.2 T, respectively. Low contrast ripples seen in **f**,**g**,**v** and **w** are from topography.

All the MFM images are scanned after the field (or temperature) is stabilized at the desired value. Even after 8 h under the same field and temperature, new COI giant stripes can still appear and some of the existing ones can grow longer as marked by the dark and white arrows, respectively, in Fig. 2l, reflecting the relaxation characteristics of the system[20]. As we continue this FR process, the COI giant stripes then fully cover the film, squeezing the FMM phase into narrow filaments (Fig. 2o, 0.5 T). When the field is removed, the FMM filaments in the COI background become broken (Fig. 2p), because the easy magnetization axis lies in the film plane (refer to ref. 30 and Supplementary Fig. 5). Note that the PS configuration in this 0 T state (Fig. 2p) is different from that of the initial 0 T state (Fig. 2a), with a vastly enhanced resistivity after cycling the field (Fig. 1b), and the observed large COI stripes accompanying the 'overshot' in resistivity is consistent with the model proposed in ref. 23.

Starting with this new initial state, we then continue to apply the field, but the direction is opposite to that of the first run (path 3, Fig. 1b). At $\mu_0 H = -0.3$ T (Fig. 2q), the broken FMM paths reconnect. From now on, the melting shows a three-step process. First, at $-1.8$ T (Fig. 2r) the COI giant stripes split into thinner ones by the formation of FMM paths from the inside, implying that at this stage the large stripes are unstable, and the split is energetically more favourable than shrinkage. Second, the COI stripes start to shrink and become thinner and thinner as the magnitude of the field increases (Fig. 2r–t). Finally, the COI filaments further shrink into segments (Fig. 2u), then to droplets (Fig. 2v) and disappear, leaving a pure FMM phase (Fig. 2w). This is similar to the first run (Fig. 2c,d) but at the same field the average stripe width is a bit larger. Again, as the field is reduced (path 4, Fig. 1b) the COI giant stripes reappear (Fig. 2x, $-1.4$ T). Interestingly, their initial growth seems to have a memory effect. For instance, those at the top left in Fig. 2y,k have similar locations and shapes, suggesting that both local pinning centres and a long-range strain pinning effect may exist[6,10]. When the sample is nearly fully covered by the COI stripes (Fig. 2a′), however, the distribution of the giant stripes is rather different (cf. Fig. 2o), indicating that the memory effect matters only in the case of low stripe density.

In the FMM-PS and frozen states, during melting the PC morphology shows stripes (Supplementary Fig. 6) and a nematic liquid-crystal-like distribution (Supplementary Fig. 5), respectively, but no COI domains reappear on reduction of the field.

The other type of COI reappearance we have probed is the TR near $T_C'$ (Fig. 1c). Field cooling at 2 T down to 10 K induces a saturated FMM phase, but it collapses when the film is subsequently warmed at 0 T to the phase boundary separating the FMM-PS and COI-PS states. This type of sharp change in $\rho(T)$ has also been observed in crystals[14], but, again, this important COI reappearance process has never been imaged before. The complete MFM images for this process along with the *in situ* resistance and relaxation measurements can be found in Supplementary Movie 1. At 100 K, the image shows that the film is totally an FMM (Fig. 3a). At 116 K (Fig. 3b–d), a small region with sharp bright and dark contrast appears, which is ascribed to a single COI domain (see Supplementary Fig. 5 for the explanation). The relaxation is greatly intensified at 118.2 K, where the COI phase starts to rapidly occupy the whole sample, while the sharp discontinuities are found ubiquitous. At higher temperatures, the FMM areas shrink into elongated (Fig. 3h–j) and then tiny isolated (Fig. 3l–n) droplets. Note that in this process the reappeared COI domains are patch like and show rather weak shape anisotropy.

Both the FR at 150 K and TR near 118 K happen in a martensitic way, but the COI domains show quite different shapes, suggesting that the PC morphology may depend on temperature. To check this possibility, the FR at 230, 190 and 130 K is examined and shown in Fig. 4a–c, respectively. At 230 K

(close to $T_{AFI} = 250$ K), the COI domains appear as small round droplets. As $H$ is further reduced, more and more droplets appear randomly, seldom growing but some end up connected as marked by the black arrows. At 0 T the morphology shows an anisotropic characteristic, as suggested by the fast Fourier transformation pattern. At 190 K, the images look similar to those at 150 K showing COI stripes, albeit with smaller domain widths ($\sim 0.27$ μm). However, at 130 K (near $T_C'$) the COI appears as irregularly shaped puddles, with pin-like FMM domains left inside to reduce the COI domains sizes and thus the accommodation strain energy. And at the moment, the movement of the COI domain walls should play a dominant role since only a few new nuclei are created. Thus, for FR the COI domains show variable-dimensional growth at different temperatures, from zero-dimensional droplets to two-dimensional puddles. However, it should be noted that the melting counterparts at these temperatures show rather similar PC behaviour (Supplementary Fig. 7).

On the basis of the above results, we can actually tune the growth of the COI domains (Fig. 4g). At 230 K, when the FR starts at 1.2 T with a few COI droplets showing up, we fix the field but decrease the temperature. Unlike the case shown in Fig. 4a where more COI droplets show up at decreasing fields, the existing droplets now start to grow along the *a* axis forming stripes. It illustrates that different PC morphologies can be obtained from the same initial state by controlling the parameters $H$ and $T$ (Fig. 4h and Supplementary Fig. 8), strongly suggesting that phase domain engineering at the microscopic level in manganite films is highly feasible.

## Discussion
Our results indicate that even in the same sample, the PC morphologies can be very different relying on the phase evolution along the different routes in the $T$–$H$ phase diagram. For this epitaxial system the doping is optimal for an FMM ground state and the anisotropic epitaxial strain is responsible for the induced COI ground state[21,30], which are mainly controlled by the two competing interactions of double exchange and Jahn–Teller distortions, respectively[24,26]. $T$ and $H$ can act as key parameters in tuning the COI and FMM phase volume fractions by cooperatively mediating the strain field, which varies with the epitaxial strain and interfacial strain of these phase domains, making the PC morphologies controllable. Moreover, as the collective Jahn–Teller distortions in the film are coupled to the NGO(001) substrate due to the interfacial octahedral connectivity[31], it is very likely that the anisotropic octahedral tilt along the *a* axis in the orthorhombic substrate can provide a preferential growth to the emergent phases, which can lead to a remarkable anisotropic distribution of the phase domains. In another strongly correlated electron system VO₂, uniaxial phase stripes were also observed in strained films through the metal–insulator transition[32]. A common feature of these systems is that the structural transformation accompanies the electronic state transition. The elongated domains along a specific direction can qualitatively account for the anisotropic transport properties in the PS regime observed for both systems[27,32], including the onset of percolation observed in this work (cf. Figs 1b and 2h,n). Specifically, from $T_{AFI}$ to $T_C'$, the COI dominates over the FMM, and the Jahn–Teller effect or the electron–lattice coupling[24] increases as $T$ decreases (Fig. 1a and Supplementary Fig. 7), while a high $H$ always favors the FMM phase due to an enhancement of the double-exchange interaction. At 230 K, the Jahn–Teller distortions are weak, only small round droplets can form during the FR. At 130 K, the Jahn–Teller distortions under the anisotropic epitaxial strain become very strong and the reappeared COI domains tend to grow very fast, leading to formation of the irregularly shaped puddles. At 190 and 150 K, the COI phase fraction ratio of the as-annealed film is moderate, and the reappeared COI domains tend to grow more compatibly along the $a$ axis to form giant stripes. Therefore, the observed PC morphology and its evolution can reflect the delicate balance of the competing interactions and their coupling to the strain field[24,26]. Nevertheless, it is still a surprise that near the FMM-PS and COI-PS phase boundary such a small variation in temperature (that is, from 150 to 130 K) can cause such a huge change in the self-organized morphology.

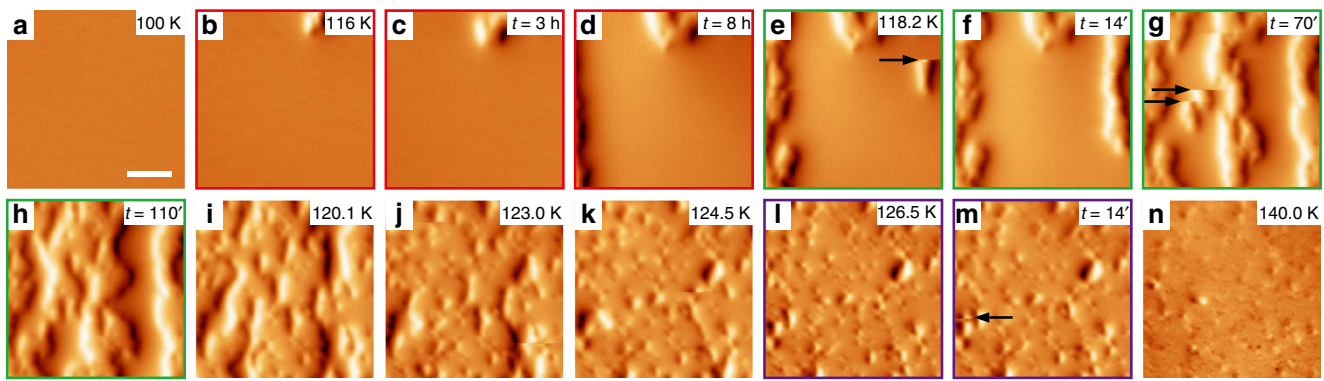

**Figure 3 | MFM image series for the TR process near $T_C'$.** (**a–n**) The images taken along the ZFW curve after cooled at 2.0 T from room temperature. A small field of 300 Oe is applied perpendicular to the film plane to enhance the magnetic signal. The scale bar is 2 μm and the scanned areas are 8 μm × 8 μm for all images. The images with the same frame colour are scanned successively at a fixed temperature but at different time as denoted. For example, **c** and **d** are taken 3 h later than **b**, respectively. The dark arrows in **e**,**g**, and **m** indicate sharp discontinuities. The colour scales for **a–n** are 1.0, 1.5, 1.5, 1.5, 2.0, 2.0, 2.5, 2.5, 2.0, 2.0, 1.5, 1.5, 1.5 and 1.0 Hz, respectively. Unlike the images taken during field sweep (under much higher fields), where only bright areas are observed for the COI domains, now the left and right edges of a certain COI domain are bright and dark, respectively, leading to a three-dimensional appearance.

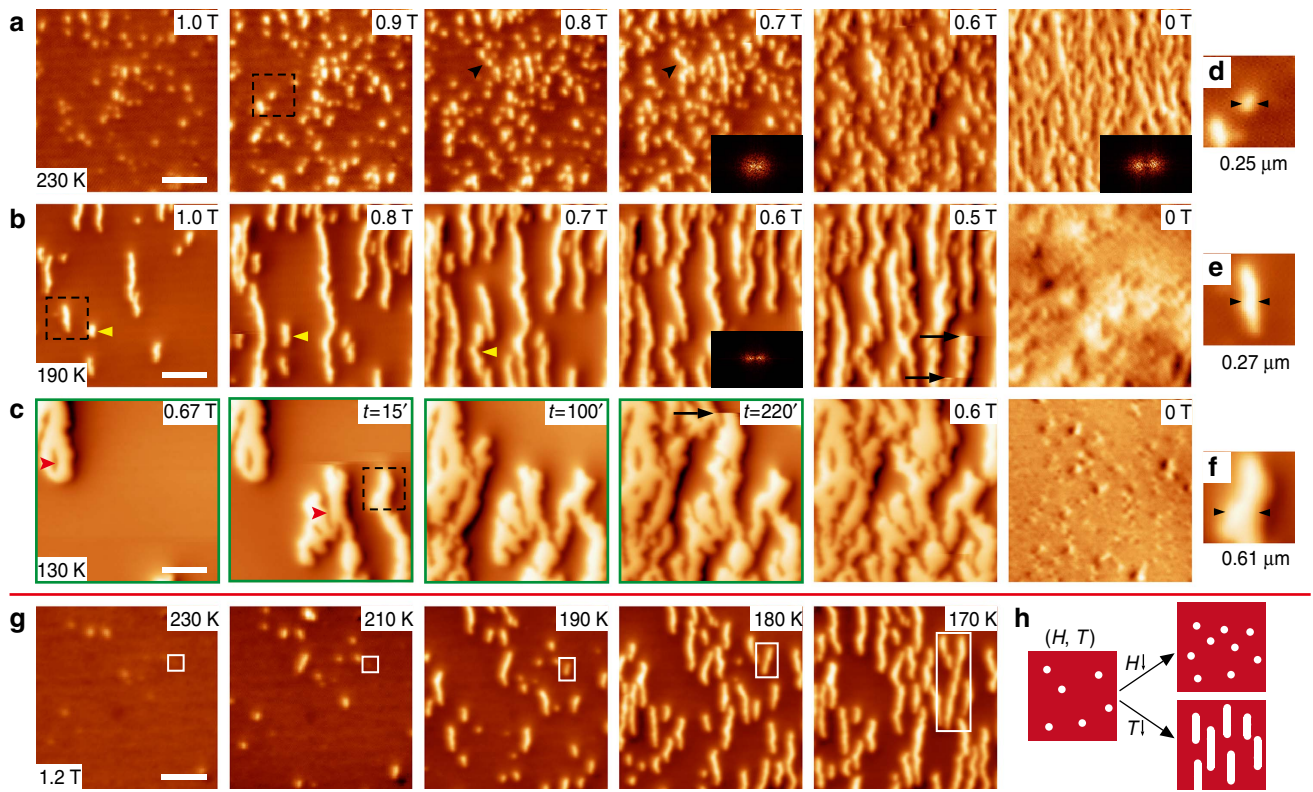

**Figure 4 | FR at different temperatures and the control of PC morphology.** (**a–c**) MFM image series taken during FR at 230, 190 and 130 K, respectively. The sample was firstly cooled from room temperature to the specified temperatures in zero field. After the field (perpendicular to the film plane) was increased to 4.0 T to fully melt the COI phase, it was then reduced. The dark arrows in **a** (0.8 and 0.7 T) indicate an example where adjacent COI droplets merge. The dark arrows in **b** (0.5 T) and **c** (0.67 T, $t = 220'$) denote sharp discontinuities. The yellow arrows in **b** (1.0, 0.8 and 0.7 T) highlight the nucleation and growth of a single COI stripe and how it merges with another one. The four images with green frame in **c** are taken under the same field but at different times as labelled. The insets in **a** and **b** are the FFT (17.0 μm$^{-1}$ × 10.5 μm$^{-1}$) for the corresponding images. (**d–f**) Enlarged views of the regions highlighted by the black dashed boxes in **a–c**, where the domain widths are measured. The irregular shapes of the COI domains in **c** make the measurement difficult but it is clear that their widths are much larger. The red arrows in **c** indicate the pin-like FMM domains. (**g**) The evolution of COI domains with decreasing temperature after they reappear at 1.2 T during FR at 230 K. The growth of a single domain is highlighted by the white boxes. (**h**) A schematic of the controlled growth of COI domains from the same initial state, by reducing the field or decreasing the temperature. All scale bars are 2 μm and all scanned areas are 8.2 μm × 8.2 μm. The colour scales are 0.3, 0.3, 0.3, 0.4, 0.5 and 0.25 Hz for **a**, 1.5, 1.5, 1.5, 2.0, 2.0 and 2.0 Hz for **b**, 4.0, 4.0, 4.0, 4.0, 4.0 and 0.5 Hz for **c**, and 0.4, 0.6, 1.2, 1.6 and 2.0 Hz for **g**, respectively.

In summary, our results provide direct and clear-cut images of the various processes along different routes in the phase diagram including the melting and in particular the FR and TR of the COI transitions, which can help understand the dynamic nature of the competing interactions in manganites. This methodology could also be applied to investigate other types of competing interactions in different types of strongly correlated electron systems. Our findings should also spark interest in applied physics as phase domains can be manipulated to exhibit different shapes and patterns with different orientations and dimensions.

## Methods

**Sample growth and transport measurements.** The 55-nm-thick LCMO films were grown on NGO(001) substrates by the pulsed laser deposition method. During growth, the oxygen pressure and temperature of the substrates were kept at 45 Pa and 735 °C, respectively, while the laser energy density and the repetition rate were set at 2 J cm$^{-2}$ and 5 Hz, respectively, resulting in a growth rate $\sim$4 mm min$^{-1}$. After deposition, the as-grown films were $ex\ situ$ annealed at 780 °C in flowing $O_2$ gas for 30 h. This $ex\ situ$ annealing process can eliminate the oxygen non-stoichiometry and significantly improve the epitaxial quality, which ensures the anisotropic strain and the interfacial octahedral coupling induced by the NGO(001) substrate. Driven by these elastic and interfacial coupling effects, both the rotations and Jahn–Teller deformations of the $MnO_6$ octahedra in the epitaxial LCMO films could be enhanced, thus leading to the localization of itinerant electrons at Mn sites and facilitating the formation of COI phase[21,30].

The films were structurally characterized by X-ray diffractions including the reciprocal space mapping using $CuK\alpha_1$ radiation (Panalytical X'pert), and the film thicknesses were determined from the Laue fringes. The magnetotransport measurements were carried out using the four-probe method on a physical property measurement system (Quantum Design) equipped with a motorized horizontal sample rotator, and the magnetic properties were measured on a vibrating sample magnetometer (Quantum Design). Detailed magnetotransport measurements of the films at various thicknesses can be found in refs 19–21,30,33.

**MFM measurements.** The MFM we used is a home-built high-field variable-temperature MFM, which is housed in a 20 T superconducting magnet (Oxford Instruments). We use a unique home-designed SpiderDrive to implement both coarse approach and image scan. The detailed description of the MFM system can be found in ref. 17. The commercial piezoresistive cantilever (PRC400 from Hitachi High-Tech Science Corporation, Japan) is used as the force sensor. The resistance of the piezoresistive cantilever is detected by a home-made preamplifier consisting of an active Wheatstone bridge and a second-stage amplifier. The resonant frequency of the cantilever is about 42 kHz. The tip is deposited with 5 nm Cr, 50 nm Fe and then 5 nm Au films using electron-beam deposition. It is magnetized perpendicular to the cantilever with a permanent magnet before it is loaded onto the scan head. The coercivity of the magnetic coating is about 250 Oe and the saturation field is about 2,000 Oe. The cantilever we used is an R9 controller with a built-in phase-locked loop from RHK Technology. The resonant frequency shift of the MFM cantilever is given as: $\frac{\delta f}{f_0} = -\frac{1}{2k}\frac{\partial F}{\partial z} \propto \frac{\partial^2 B_z}{\partial z^2}$, where $f_0$, $F$ and $k$ denote the resonant frequency, the force exerted on the tip by the sample and the spring constant of the cantilever, respectively[34]. This implies that the MFM is only sensitive to the out-of-plane component of the stray field ($B_z$).

The MFM images are collected in constant height mode. First, the topography (Supplementary Fig. 3) is imaged using contact mode and the slopes along the fast and slow scan directions are compensated. The tip is then lifted to a height of roughly 100 nm from the sample surface. For each set of MFM images, the lift height is the same but between different data sets it varies a little. The frequency-modulation mode is used to image the magnetic signals. The scan time is 0.8 s. The slow scan direction is always from top to bottom. If not otherwise specified, the fast scan direction for the images presented in this paper is from right to left, which is the normal direction. The images scanned in the opposite direction are also collected, which are very similar to the images scanned in the normal direction except that a very small shift exists in the fast scan direction due to the hysteresis of the piezo material (piezo tube scanner).

In a high field, the magnetic moments in the FMM phase are aligned with the external field. The magnetic force on the tip from an FMM domain is attractive (resonant frequency shift $\delta f < 0$) and there is no force between a COI domain and the tip ($\delta f = 0$), thus a COI domain in an FMM background shows bright. When the field is very low, the magnetic moments in the FMM domains are mostly in plane ($b$ axis is the easy magnetization axis) and the force gradient in $z$ direction is small (the $z$ component of the field, $B_z \approx 0$) except near the phase domain boundaries (Supplementary Fig. 5). Thus, the central area of an FMM or COI domain shows the same contrast ($\delta f = 0$, grey), while the opposite sides (along $b$ axis) of an FMM or COI domain show opposite contrasts: bright ($\delta f > 0$) and dark ($\delta f < 0$).

The above contrast analysis can also be supported as follows: (1) for Fig. 2d–g, on increasing the magnetic field, the overall dark area increases at the expense of the overall bright area, and it always corresponds to a large drop in resistivity (path 1 in Fig. 1b). If the dark and bright regions were both FMM domains with different magnetization orientations, the movement and annihilation of the magnetic domain walls would not result in such a large drop in resistivity; (2) if the bright regions in Fig. 2h–o were the FMM phase with magnetization antiparallel to the magnetization of the tip, they should reappear at 10 K when the field is decreased (similar to the case at 150 K). This is contrary to our observations (Supplementary Fig. 5).

## References

1. Dagotto, E., Hotta, T. & Moreo, A. Colossal magnetoresistant materials: the key role of phase separation. *Phys. Rep.* **344**, 1–153 (2001).
2. Dagotto, E. Complexity in strongly correlated electronic systems. *Science* **309**, 257–262 (2005).
3. Rini, M. *et al.* Control of the electronic phase of a manganite by mode-selective vibrational excitation. *Nature* **449**, 72–74 (2007).
4. Li, T. *et al.* Femtosecond switching of magnetism via strongly correlated spin-charge quantum excitations. *Nature* **496**, 69–73 (2013).
5. Zhang, L., Israel, C., Biswas, A., Greene, R. L. & de Lozanne, A. Direct observation of percolation in a manganite thin film. *Science* **298**, 805–807 (2002).
6. Sarma, D. D. *et al.* Direct observation of large electronic domains with memory effect in doped manganites. *Phys. Rev. Lett.* **93**, 097202 (2004).
7. Wu, W. *et al.* Magnetic imaging of a supercooling glass transition in a weakly disordered ferromagnet. *Nat. Mater.* **5**, 881–886 (2006).
8. Israel, C. *et al.* Translating reproducible phase-separated texture in manganites into reproducible two-state low-field magnetoresistance: An imaging and transport study. *Phys. Rev. B* **78**, 054409 (2008).
9. Murakami, Y. *et al.* Ferromagnetic domain nucleation and growth in colossal magnetoresistive manganite. *Nat. Nanotechnol.* **5**, 37–41 (2010).
10. Lai, K. *et al.* Mesoscopic percolating resistance network in a strained manganite thin film. *Science* **329**, 190–193 (2010).
11. Burkhardt, M. H. *et al.* Imaging the first-order magnetic transition in $La_{0.35}Pr_{0.275}Ca_{0.375}MnO_3$. *Phys. Rev. Lett.* **108**, 237202 (2012).
12. Rawat, R., Kushwaha, P., Mishra, D. K. & Sathe, V. G. Direct visualization of first-order magnetic transition in $La_{5/8-y}Pr_yCa_{3/8}MnO_3$ ($y = 0.45$) thin films. *Phys. Rev. B* **87**, 064412 (2013).
13. Du, K. *et al.* Visualization of a ferromagnetic metallic edge state in manganite strips. *Nat. Commun.* **6**, 6179 (2015).
14. Kuwahara, H., Tomioka, Y., Asamitsu, A., Moritomo, Y. & Tokura, Y. A first-order phase transition induced by a magnetic field. *Science* **270**, 961–963 (1995).
15. Tomioka, Y., Asamitsu, A., Moritomo, Y., Kuwahara, H. & Tokura, Y. Collapse of a charge-ordered state under a magnetic field in $Pr_{1/2}Sr_{1/2}MnO_3$. *Phys. Rev. Lett.* **74**, 5108–5111 (1995).
16. Uehara, M., Mori, S., Chen, C. H. & Cheong, S. W. Percolative phase separation underlies colossal magnetoresistance in mixed-valent manganites. *Nature* **399**, 560–563 (1999).
17. Zhou, H., Wang, Z., Hou, Y. & Lu, Q. A compact high field magnetic force microscope. *Ultramicroscopy* **147**, 133–136 (2014).
18. Biswas, A. *et al.* Strain-driven charge-ordered state in $La_{0.67}Ca_{0.33}MnO_3$. *Phys. Rev. B* **63**, 184424 (2001).
19. Huang, Z. *et al.* Phase evolution and the multiple metal-insulator transitions in epitaxially shear-strained $La_{0.67}Ca_{0.33}MnO_3/NdGaO_3$(001) films. *J. Appl. Phys.* **108**, 083912 (2010).
20. Huang, Z. *et al.* Dynamic phase separation as revealed by the strong resistance relaxation in epitaxially shear-strained $La_{0.67}Ca_{0.33}MnO_3/NdGaO_3$(001) thin-films. *J. Magn. Magn. Mater.* **322**, 3544–3550 (2010).
21. Wang, L. F. *et al.* Annealing assisted substrate coherency and high-temperature antiferromagnetic insulating transition in epitaxial $La_{0.67}Ca_{0.33}MnO_3/NdGaO_3$(001) films. *AIP Adv.* **3**, 052106 (2013).
22. García-Muñoz, J. L., Collado, A., Aranda, M. A. G. & Ritter, C. Multilevel hierarchy of phase separation processes in $La_{5/8-y}Pr_yCa_{3/8}MnO_3$. *Phys. Rev. B* **84**, 024425 (2011).

23. Khomskii, D. & Khomskii, L. Fine mist versus large droplets in phase separated manganites. *Phys. Rev. B* **67,** 052406 (2003).

24. Mathur, N. & Littlewood, P. Mesoscopic texture in manganites. *Phys. Today* **56,** 25–30 (2003).

25. Podzorov, V., Kim, B. G., Kiryukhin, V., Gershenson, M. E. & Cheong, S. W. Martensitic accommodation strain and the metal-insulator transition in manganites. *Phys. Rev. B* **64,** 140406 (2001).

26. Mathur, N. D. & Littlewood, P. B. The self-organised phases of manganites. *Solid State Commun.* **119,** 271–280 (2001).

27. Ward, T. Z. *et al.* Elastically driven anisotropic percolation in electronic phase-separated manganites. *Nat. Phys.* **5,** 885–888 (2009).

28. Sharma, P. A. *et al.* Phase-segregated glass formation linked to freezing of structural interface motion. *Phys. Rev. B* **78,** 134205 (2008).

29. Sharma, P. A., Kim, S. B., Koo, T. Y., Guha, S. & Cheong, S. W. Reentrant charge ordering transition in the manganites as experimental evidence for a strain glass. *Phys. Rev. B* **71,** 224416 (2005).

30. Huang, Z. *et al.* Tuning the ground state of $La_{0.67}Ca_{0.33}MnO_3$ films via coherent growth on orthorhombic $NdGaO_3$ substrates with different orientations. *Phys. Rev. B* **86,** 014410 (2012).

31. Rondinelli, J. M., May, S. J. & Freeland, J. W. Control of octahedral connectivity in perovskite oxide heterostructures: an emerging route to multifunctional materials discovery. *MRS Bulletin* **37,** 261–270 (2012).

32. Liu, M. K. *et al.* Anisotropic electronic state via spontaneous phase separation in strained vanadium dioxide films. *Phys. Rev. Lett.* **111,** 096602 (2013).

33. Wang, L. F. *et al.* Pseudomorphic strain induced strong anisotropic magnetoresistance over a wide temperature range in epitaxial $La_{0.67}Ca_{0.33}MnO_3/NdGaO_3$(001) films. *Appl. Phys. Lett.* **97,** 242507 (2010).

34. Hartmann, U. Magnetic Force Microscopy. *Annu. Rev. Mater. Sci.* **29,** 53–87 (1999).

## Acknowledgements

We gratefully thank Professor Alex de Lozanne for his suggestions, and Feng Jin and Qiyuan Feng for their help in the experiments. This work is supported by National Natural Science Foundation of China (Grant Nos 11274287, 11204306, U1232210, 11374278, U1432251 and 11474263), the National Basic Research Program of China (Grant Nos 2012CB927402 and 2015CB921201), the project of Chinese national high magnetic field facilities, and the facilities of the National Synchrotron Radiation Laboratory, University of Science and Technology of China.

## Author contributions

Q.L. and W.W. designed the experiments. L.W. and Z.H. grew the manganite film and performed the transport measurements. H.Z. and Y.H. conducted the MFM measurements. H.Z., Q.L. and W.W. analysed the MFM data and wrote the manuscript. All authors participated in the discussion during the preparation of the manuscript.

## Additional information

**Competing financial interests:** The authors declare no competing financial interests.